\documentclass[10pt,twocolumn,prd,amssymb,nofootinbib]{revtex4}
\usepackage{amsmath, amsthm, amscd, amssymb}
\usepackage{graphicx}

\usepackage[format=hang,font=small,labelfont=bf]{caption}
\usepackage[colorlinks=true,citecolor=red,linkcolor=blue,urlcolor=blue]{hyperref}

\begin{document}

\title{Variable Brane Tension and Dark Energy}
\author{Naman Kumar\footnote{The author's affiliation has changed since the original submission. The current affiliation is Department of Physics, Indian Institute of Technology Gandhinagar, Gujrat, India, 382355}}
\email{namankumar5954@gmail.com}
\affiliation{School of Basic Sciences, Indian Institute of Technology, Bhubaneswar, India, 752050 }

\begin{abstract}
In this letter, we show that in a particular braneworld scenario with variable brane tension, we obtain matter acting as dark energy while the gravitational constant $G$ promoted to a scalar field on the brane plays the role of matter (both in the sense they have an 'effective' Equation of State equivalent to that of dark energy and matter respectively). This result is interpreted from the Friedmann equation obtained from our model that exactly matches the standard Friedmann equation of General Relativity with a cosmological constant $\Lambda$ in terms of the aforementioned quantities. The universe is assumed to consist of only matter and dark energy in this model which is a good approximation for our universe.
\end{abstract}
\maketitle
\section{Introduction}
The problem of dark energy is one of the most challenging problems in cosmology. The need for dark energy emerged out of necessity to explain the observation that the universe is expanding in an accelerated manner\cite{riess1998observational}. It has remained a mystery since and is yet to be solved. In the framework of General Relativity, a small and positive cosmological constant $\Lambda$\cite{carroll2001cosmological} is considered as dark energy and since it is proportional to the metric $g_{\mu\nu}$, it is interpreted as vacuum energy. However, its true nature remains elusive since it has an equation of state $\omega=-1$ and also violates the strong energy condition $(T_{\mu\nu}-\frac{1}{2}Tg_{\mu\nu})u^\mu u^\nu<0$. A number of alternative approaches have been proposed such as quintessence field\cite{tsujikawa2013quintessence}, modified gravity theories(see\cite{shankaranarayanan2022modified} for a discussion), and braneworld cosmology(see\cite{brax2003cosmology} for a review). In the braneworld scenario, for instance, it was shown in\cite{dvali20004d} by considering the brane in 5D Minkowski bulk, that the Newtonian potential is that of 5D theory at large distances explaining dark energy\cite{deffayet2002accelerated}. This theory effectively has a 4D massless graviton plus a scalar field as its tensorial structure. Recently, it was shown in\cite{garcia2018brane} that a variable brane tension can explain dark energy where the brane tension was proposed to depend upon the scale factor. There has also been a growing theoretical effort to apply the holographic principle to the dark energy problem called the holographic dark energy paradigm (see\cite{wang2017holographic} for a review). It has led to Kaniadakis holographic dark energy\cite{drepanou2022kaniadakis} (which is based on Kaniadakis entropy and is a generalization of Boltzmann-Gibbs entropy), Barrow holographic dark energy\cite{saridakis2020barrow} (which uses Barrow entropy instead of the standard Bekenstein-Hawking entropy) and Power-law holographic dark energy\cite{telali2022power}.\\
In this paper, we use the braneworld model with a variable brane tension to explain the problem of dark energy. It is shown that the matter effectively acts as the dark energy and the 4D Newton's constant $G$ promoted to a scalar field on the brane plays the role of matter. The idea is elaborated in the next section.

\section{Emergence of Dark Energy}
For a constant brane tension, Einstein's equation on the brane (with no cosmological constant) reads
\begin{equation}
G_{\mu\nu}=\kappa_{(4)}^2T^m_{\mu\nu}+\frac{\kappa_{(4)}^2}{\lambda}S_{\mu\nu}-\xi_{\mu\nu}+\sqrt{\frac{\kappa_{(4)}^2}
{\lambda}}F_{\mu\nu}
\end{equation}
Here, $T_{\mu\nu}^m$ is the matter energy-momentum tensor, $S_{\mu\nu}$ is the quadratic energy-momentum tensor, $\xi_{\mu\nu}$ is the non-local Weyl tensor and $F_{\mu\nu}$ is the matter field in the bulk. $\kappa_{(4)}^2=8\pi G_N=\kappa_{(5)}^4\lambda$ is the constant with $\lambda$ being the constant brane tension. Let us assume that the bulk is empty so that $F_{\mu\nu}=0$ and therefore the pullback on the brane is also zero, which is related to the non-standard model fields\cite{wong2010inflation} as well as the bulk is $AdS_5$ such that the Weyl term $\xi_{\mu\nu}$ vanishes. This gives 
\begin{equation}
   G_{\mu\nu}=\kappa_{(4)}^2T^m_{\mu\nu}+\frac{\kappa_{(4)}^2}{\lambda}S_{\mu\nu} \label{cbtfe}
\end{equation}
Let us now consider the case where the gravitational constant is promoted to a scalar field $G$ such that $\kappa_{(4)}^2=8\pi G$ and the five-dimensional gravitational constant is related as usual as $k^4_{(5)}\lambda=8\pi G$. In this case, the field equation becomes
\begin{equation}
    G_{\mu\nu}=k_{(5)}^4\lambda( T^M_{\mu\nu}+T^G_{\mu\nu})+k_{(5)}^4S_{\mu\nu}\label{vbtfe}
\end{equation}
where $T^{\mu\nu}_G$ is the energy-momentum tensor of the scalar field $G$ on the brane given as
\begin{equation}
    T^G_{\mu\nu}=(\partial_\mu G)(\partial_\nu G)-g_{\mu\nu}\bigg[\frac{1}{2}(\partial_\sigma G)(\partial^\sigma G)-V(G)\bigg]
\end{equation}
Note that since the brane tension is now a variable and the energy-momentum tensor of matter and scalar field is divergenceless, Einstein's equation requires the following relation to hold
\begin{equation}
    k_{(5)}^4T^{\mu\nu}\nabla_\mu\lambda+k_{(5)}^4\nabla_\mu S^{\mu\nu}-\nabla_\mu \xi^{\mu\nu}=0
\end{equation}
where $T^{\mu\nu}$ is the energy-momentum tensor of matter plus the scalar field $G$. The field equation (\ref{vbtfe}) can be derived from the action $S=S_{bulk}+S_{brane}$ on the brane where
\begin{align}
    S_{bulk}=\frac{1}{2k_{(5)}^2}\int d^5x\sqrt{-g^{(5)}}R^{(5)}\\
    S_{brane}=\int d^4x\sqrt{-g^{(4)}}\bigg[\bigg(-\frac{1}{2}g^{\mu\nu}\partial_\mu G\partial_\nu G-V(G)\bigg)+\mathcal{L}_M\bigg]\label{brane_actiom}
\end{align}
where $\mathcal{L}_M$ is the matter Lagrangian on the brane. Considering the flat FLRW geometry of the brane, the Friedmann equation on the brane reads
\begin{align}
    3H^2=k_{(5)}^4\lambda\sum_i\rho_i\bigg(1+\frac{\rho_i}{2\lambda}\bigg)\label{friedmann2}
\end{align}
where $\Sigma \rho_i=\rho_M+\rho_r+\rho_G$ is the total energy density(of matter, radiation, and scalar field $G$ respectively). Since we are interested in late-time cosmology, the second term in the brackets of (\ref{friedmann2}) drops out in the low energy limit and we can write the above Friedmann equation in terms of redshift factor $z$ as
\begin{equation}
    H^2=H_0^2[\lambda(\Omega_{M0}(1+z)^3+\Omega_{G0}(1+z)^{3(1+w)}+\Omega_{r0}(1+z)^4)]\label{friedmann}
\end{equation}
where the density parameter $\Omega_i(t)$ is given by
\begin{equation}
   \Omega_i(t)=\frac{k^4_{(5)}}{3H^2(t)}\rho_i(t) \label{density_parameter}
\end{equation}
and is the same as that in General Relativity in the sense that they carry the same information. $w$ is the Equation of State(EoS) of the scalar field $G$. For the scalar field(assuming no spatial variation in $G$), we have
\begin{align}
    \rho_G=\frac{\dot G^2}{2}+V(G)\label{rhoG}\\
    p_G=\frac{\dot G^2}{2}-V(G)\label{pG}
\end{align}
If we now consider the kinetic term to be much greater than the potential term $\dot G^2>>V(G)$ which can be thought of as the scalar field $G$ rolling down a very steep potential, then we have $w=1$. Using the divergenceless property of the scalar field energy-momentum tensor($\nabla_\mu T^{\mu\nu}_G=0$) and equations (\ref{rhoG}) and (\ref{pG}), the Equation of Motion(EoM) of the scalar field coupled to gravity is given by(assuming no spatial variation in $G$)\footnote{The EoM of the scalar field can also be obtained by extremizing the action(\ref{brane_actiom}).}
\begin{equation}
    \ddot G+3H\dot G+\frac{dV(G)}{dG}=0
\end{equation}
Thus, $w=1$ means that the friction term due to the Hubble parameter $H$ is non-dominant and the scalar field easily rolls down the potential generating a larger kinetic term. The (variable) brane tension $\lambda$ is taken to be the function of the redshift factor $z$ as
\begin{equation}
    \lambda(z)=\lambda_0(1+z)^{-3}
\end{equation}
Inserting this value of $\lambda$ and $w=1$ in (\ref{friedmann}) with $\Omega_{r0}=0$ for our universe, we obtain
\begin{equation}
   H^2=H_0^2[\lambda_0(\Omega_{M0}+\Omega_{G0}(1+z)^3)] \label{friedmann3}
\end{equation}
This is equivalent to the standard Friedmann equation of General Relativity with cosmological constant $\Lambda$ (and $\Omega_{r0}=0$) such that
\begin{align}
    \lambda_0\Omega_{M0}=\Omega_{\Lambda0}\approx0.7\\
    \lambda_0\Omega_{G0}=\Omega_{mo}\approx0.3
\end{align}
Therefore, in this braneworld model, the scalar field $G$ effectively acts as the matter in the sense that although its EoS is w=1, the 'effective' EoS (i.e., after taking the effect of the variable brane tension) is that of matter. This can be interpreted from the Friedmann equation(\ref{friedmann3}) while the matter field effectively acts as the cosmological constant! This result is quite surprising.
\section{Conclusion}
In this letter, we presented a possible explanation of the dark energy problem by using the braneworld model with a variable brane tension. The four-dimensional Newton's constant is promoted to a scalar field on the brane. Counterintuitively, it turned out that the matter on the brane acts as an effective dark energy term, while the scalar field on the brane effectively acts as matter. Considering the late-time cosmology, the Friedmann equation obtained using this model exactly matches the standard Friedmann model of General Relativity with a cosmological constant and $\Omega_{r0}=0$ for our universe. Therefore, in this particular braneworld scenario, we successfully achieved our aim of explaining dark energy without the need to introduce any extra field.
\newpage
\section{Addendum}
In earlier work, we used a variable brane tension (VBT) framework to explain the late-time cosmic acceleration of the universe. The idea is to promote the 4D Newton's constant $G$ on the brane to a scalar field $G(\phi)$. The 5D Newton's constant is held fixed by subsequently varying the brane tension $\lambda$. This gives the Friedmann equation on the brane as
\begin{equation}
  H^2=H_0^2[\lambda_0(\Omega_{M0}+\Omega_{G0}(1+z)^3)]  \label{friedmannn}
\end{equation}
where $H=\dot a/a$ ($a$ is the scale factor), $z$ is the redshift, $\lambda_0$ is a constant, and the last term corresponds to the scalar field $G$. (\ref{friedmannn}) shows that matter "effectively" acts as dark energy while the scalar field plays the role of matter on the brane such that 
\begin{align}
    \lambda_0\Omega_{M0}\approx0.7\\
    \lambda_0\Omega_{G0}\approx0.3
\end{align}
The variable brane tension was taken to be $\lambda=\lambda_0a^3$. Let us extend this to a more general form 
\begin{equation}
    \lambda=\lambda_0\bigg[1+\bigg(\frac{T_c}{T}\bigg)^n\bigg]
\end{equation}
When the temperature $T$ is large compared to a critical value $T_c$, i.e., $T>>T_c$, $\lambda$ behaves as a constant but when $T_c>>T$, then $\lambda$ varies with temperature $T$. Also, we use the standard result that $T\propto a^{-1}$. Therefore, for the universe at late times, $T_c>>T$ and $\lambda\propto T^{-n}=a^n$. In the previous work, it was shown that $n=3$. Let us write the general form of the Friedmann equation on the brane with a VBT (assuming $AdS_5$ bulk and no matter fields in the bulk) which reads\cite{gergely2008friedmann,wong2010inflation,garcia2018brane}
\begin{equation}
    3\bigg(\frac{\dot a}{a}\bigg)^2=k_{(5)}^4\lambda\sum\rho_i\bigg(1+\frac{\rho_i}{2\lambda}\bigg)\label{vbt_friedmann}
\end{equation}
where $\sum\rho_i=\rho_M+\rho_r+\rho_\phi$. The pressure ($p$) and density ($\rho$) of the scalar field $G(\phi)$ is given as
\begin{align}
    \rho_\phi=\frac{\dot\phi^2}{2}+V(\phi)\\
    p_\phi=\frac{\dot\phi^2}{2}-V(\phi)
\end{align}
In early times, the temperature of the universe was very high such that $T>>T_c$; therefore, $\lambda=\lambda_0$ is a constant. Therefore, using (\ref{vbt_friedmann}), we get
\begin{equation}
    \frac{\ddot a}{a}=-\frac{k_{(5)}^4\rho_i}{6}\bigg(2\rho_i+3p\bigg)\label{vbt_second_friedmann}
\end{equation}
Since for the early universe, $\rho^2$ would dominate, we have ignored the first term in (\ref{vbt_friedmann}) to reach (\ref{vbt_second_friedmann}). For accelerated expansion, the pressure $p$ must be negative. This requires that for inflation 
\begin{equation}
    V(\phi)>>\dot\phi^2
\end{equation}
For our purpose, we need a potential such that the field $\phi$ rolls slowly, maintaining the slow-roll conditions, and then it keeps rolling, gaining kinetic energy that subsequently dominates the potential. Thus, the field $\phi$ starts the inflation, and once slow-roll conditions are not satisfied, inflation stops. $\phi$ keeps rolling down and now satisfies $\dot\phi^2>>V(\phi)$ which works at dark energy at late times. A schematic diagram is shown below. This corresponds to 
\begin{equation}
    V(\phi)=e^{-\alpha\phi^2}
\end{equation}
where $\alpha>0$.
\begin{figure}[ht!]
    \centering
    \includegraphics[width=\linewidth]{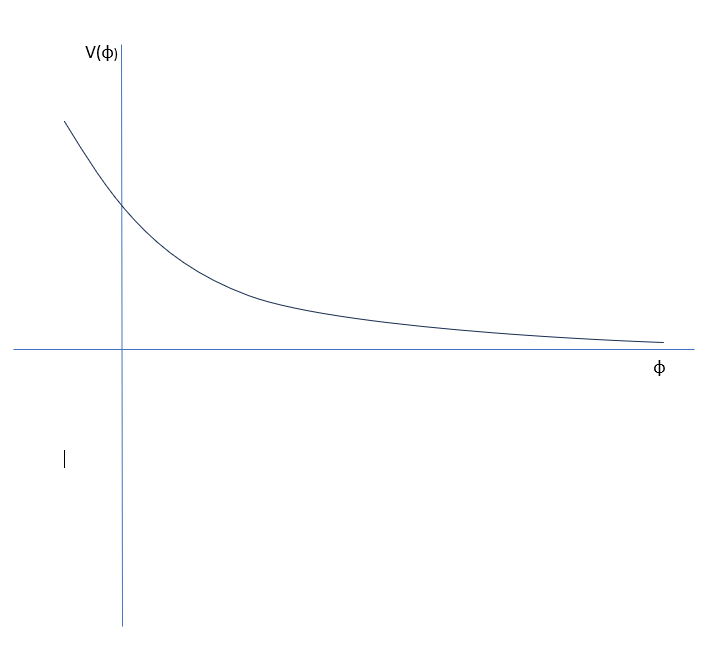}
    \caption{The field $\phi$ rolls down slowly, maintaining the slow-roll conditions, and gradually satisfies $\dot\phi^2>>V(\phi)$ throughout. This behaviour later contributes to late time cosmic acceleration.}
    \label{fig1}
\end{figure}
\newpage
However, there is one problem: reheating. If $\phi$ does not settle to a stable minimum and oscillate about it, it cannot participate in reheating, which is a crucial aspect post-inflation. For this, we promote $G$ to a complex scalar field $G(\Phi)$. This is equivalent to two scalar fields $G\to G(\phi,\chi)$ such that $\Phi=\phi+i\chi$. The imaginary part of the complex scalar field $\chi$ participates in reheating while the real part $\phi$ behaves as inflation and generates a scale-invariant power spectrum by satisfying slow-roll conditions consistent with CMB observations. The effective Lagrangian can be written as
\begin{equation}
    \mathcal{L}=\frac{1}{2}g^{\mu\nu}\partial_\mu\phi\partial_\nu\phi+\frac{1}{2}g^{\mu\nu}\partial_\mu\chi\partial_\nu\chi-V(\phi,\chi)
\end{equation}
where
\begin{equation}
    V(\phi,\chi)=e^{-\alpha\phi^2}+\frac{1}{2}m_2^2\chi^2
\end{equation}
So, this effectively becomes two-field inflation, one driving inflation while the other leading to reheating since $\chi$ has a stable minimum at $\chi=0$ about which it oscillates post-inflation, decaying into Standard Model (SM) particles. $\chi$ can also contribute to Primordial Black Holes(PBHs) formation if it satisfies ultra-slow roll conditions (see\cite{garcia2017primordial,germani2017primordial}) before reaching its minimum. The advantage of this two-field inflation is that $\phi$ independently drives inflation and satisfies slow-roll conditions to give the correct power spectrum while $\chi$ independently satisfies some ultra-slow roll condition before reaching its minimum and thus participates in both reheating and PBHs formation by generating a spiked power spectrum for the time it satisfies the ultra slow-roll conditions. Note that $\chi$ has to satisfy some ultra slow-roll conditions; otherwise, its density would be swept away by inflation, rendering it impossible to be able to participate in reheating. Therefore, in this two-field inflation scenario, one need not worry that in satisfying the ultra-slow roll conditions, the universe may exit the inflationary phase early, as may happen in the single-field inflation case. Furthermore, after reheating, $\chi$ has decayed into SM  particles, and $G$ becomes a single field $G(\phi)$ resulting in the effective Lagrangian given by
\begin{equation}
  \mathcal{L}=\frac{1}{2}g^{\mu\nu}\partial_\mu\phi\partial_\nu\phi-V(\phi)  
\end{equation}\\
which is consistent with Lagrangian on the brane at late times, consistent with our previous work.\\
To summarize, the scalar field $\phi$ drives inflation, which ceases when slow-roll conditions are not satisfied. The field $\phi$ then subsequently rolls, satisfying the new condition $\dot\phi^2>>V(\phi)$, which leads to late-time cosmic acceleration. Another scalar field $\chi$ oscillates about its minimum at the end of inflation, leading to reheating. It can also lead to the formation of PBHs by satisfying ultra-slow roll conditions for an appropriate time period before reaching its minimum. After reheating, $G$ effectively behaves as a single scalar field $G(\phi)$. Both the scalar fields are part of a single complex scalar field $\Phi$. Therefore, the braneworld scenario(BWS) provides a beautiful transition from inflation to late-time cosmic acceleration since field $\phi$, which initially satisfies the Equation of state(EoS) $w=-1$, leading to inflation, will eventually satisfy the EoS $w=1$ since it rolls down and gain kinetic energy. Usually, $w=-1$ is required to explain dark energy too, but as shown in the previous work in the BWS, dark energy is explained for $w=1$.

\bibliography{bib}
\bibliographystyle{unsrt}
\end{document}